%% file: main.tex
\documentclass[conference,final]{IEEEtran}
\IEEEoverridecommandlockouts

\usepackage[english]{babel}
\usepackage{cite}
\usepackage{amsmath,amssymb,amsfonts}
\usepackage{algorithmic}
\usepackage{graphicx}
\usepackage{textcomp}
\usepackage{xcolor}
\usepackage{multirow}
\usepackage{makecell}
\usepackage{booktabs}
\usepackage{hyperref}
\usepackage{amsmath}
\usepackage[numbers]{natbib}

\newcolumntype{P}[1]{>{\centering\arraybackslash}p{#1}}
\renewcommand{\arraystretch}{1.2}
\newcommand{\our}{{AMCAD}}

\begin{document}

\title{
\our{}: Adaptive Mixed-Curvature Representation based Advertisement Retrieval System
}

\author{
\IEEEauthorblockN{
Zhirong Xu$^{\ast \dagger}$\thanks{$^{\dagger}$Both authors contributed equally to this work.}~~~~~Shiyang Wen$^{\ast \dagger}$~~~~~Junshan Wang$^{\ast}$~~~~~Guojun Liu$^{\ast}$~~~~~Liang Wang$^{\ast}$~~~~~Zhi Yang$^{\ddagger}$~~~~~\\Lei Ding$^{\ast}$~~~~~Yan Zhang$^{\ast}$~~~~~Di Zhang$^{\ast}$~~~~~Jian Xu$^{\ast}$~~~~~Bo Zheng$^{\ast}$
}
\textit{$^\ast$Alibaba Group}
\textit{$^\ddagger$Peking University}\\
Beijing, China\\
\textit{$^\ast$\{zhirong.xzr, shiyang.wsy, junshan.wjs, guojun.liugj, liangbo.wl, ziying.dl, zy143424\}@alibaba-inc.com}\\
\textit{$^\ast$\{di.zhangd, xiyu.xj, bozheng\}@alibaba-inc.com}
\textit{$^\ddagger$yangzhi@pku.edu.cn}\\
}

\maketitle

\begin{abstract}
Graph embedding based retrieval has become one of the most popular techniques in the information retrieval community and search engine industry.
The classical paradigm mainly relies on the flat Euclidean geometry.
In recent years, hyperbolic (negative curvature) and spherical (positive curvature) representation methods have shown their superiority to capture hierarchical and cyclic data structures respectively. 
However, in industrial scenarios such as e-commerce sponsored search platforms, the large-scale heterogeneous query-item-advertisement interaction graphs often have multiple structures coexisting.
Existing methods either only consider a single geometry space, or combine several spaces manually, which are incapable and inflexible to model the complexity and heterogeneity in the real scenario.
To tackle this challenge, we present a web-scale Adaptive Mixed-Curvature ADvertisement retrieval system (\our{}) to automatically capture the complex and heterogeneous graph structures in non-Euclidean spaces.
Specifically, entities are represented in adaptive mixed-curvature spaces, where the types and curvatures of the subspaces are trained to be optimal combinations.
Besides, an attentive edge-wise space projector is designed to model the 
similarities between heterogeneous nodes according to local graph structures and the relation types.
Moreover, to deploy \our{} in Taobao, one of the largest e-commerce platforms with hundreds of million users, we design an efficient two-layer online retrieval framework for the task of graph based advertisement retrieval.
Extensive evaluations on real-world datasets and A/B tests on online traffic are conducted to illustrate the effectiveness of the proposed system. 
\end{abstract}


\input{introduction.tex}

\input{problem.tex}

\input{preliminary.tex}

\input{model.tex}

\input{implementation.tex}

\input{experiments.tex}

\input{related.tex}

\input{conclusion.tex}

\bibliographystyle{IEEEtranN}
\bibliography{main}

\end{document}

%% file: introduction.tex
\section{Introduction}

Sponsored search (SS) is a multi-billion dollar business model which has been widely used in e-commerce platforms. In general, the users express search intent by posing queries, and clicking/purchasing the sponsored products (ads) and organic products (equal to items in this paper) upon interests. These search behaviors link the queries, ads and items into a large-scale heterogeneous graph. The ad retrieval system of e-commerce platforms aims to retrieve a set of ads catching the users' interest. This goal aligns well with the link prediction problem on this heterogeneous graph, i.e., to find a set of ads highly probable to be linked to the queries.

Graph representation learning, especially Graph Neural Networks (GNNs) \cite{ying2018graph,velivckovic2017graph}, has received significant interest in recent years. Due to the state-of-the-art performance in many graph-based tasks such as link prediction, graph embedding based retrieval (GEBR) has been an active research topic in the information retrieval community and search engine industry \cite{fan2019mobius,ying2018graph,pfadler2020billion,xie2021sequential,li2020hierarchical,zheng2020price}. In this paper, we consider the problem of developing and deploying GEBR system in the context of a real industry setting, Taobao\footnote{\href{https://www.taobao.com/}{https://www.taobao.com/}}, the largest e-commerce platform in China. Different from academic studies on small datasets, GEBR is a challenging problem in industrial search engines due to the complex, large-scale query-item-ad interaction graph.

\begin{figure}
  \centering
  \includegraphics[width=\linewidth]{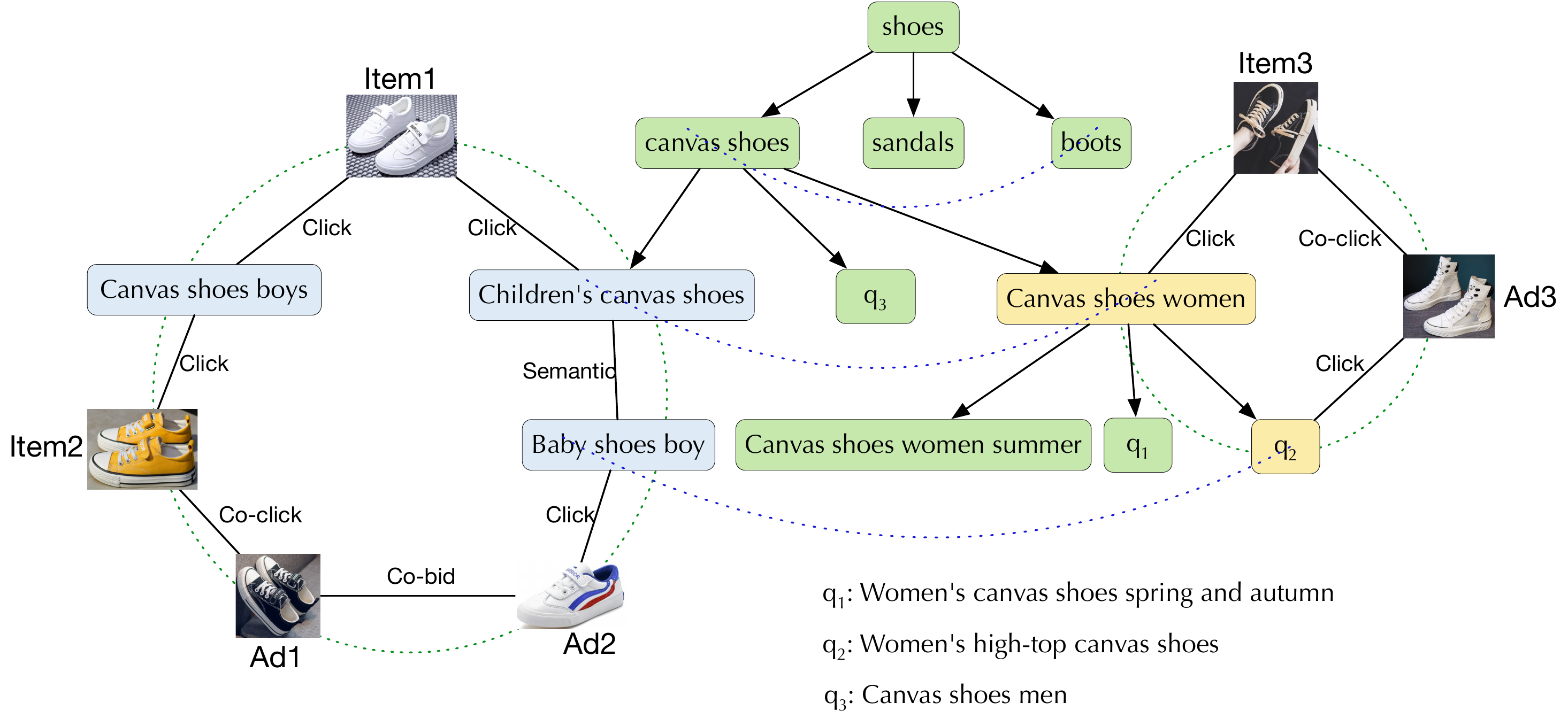}
  \caption{The complex and heterogeneous structure in the interaction graph of e-commerce search, which demonstrates two mixed structures: tree and cyclic. Note that the same node can be in different structures simultaneously.}
  \label{fig:tree_cyclic_structure_sample}
\end{figure}

In Taobao, the graph data are complex with varied structures. As illustrated in Fig.~\ref{fig:tree_cyclic_structure_sample}, since each query can be mapped to a node in Taobao's category tree, the queries can be organized as a tree intrinsically. Moreover, The superordinate, hyponym and subordinate relationships designate a more fine-grained hierarchical structure of queries. Besides the tree structure, complex relationships between different items and ads, such as co-clicks and semantic similarities, tend to be formed as a cyclic structure. In such cases, Euclidean space approaches suffer from large distortion when representing these structures \cite{sala2018representation,bachmann2020constant}. Recently it has been shown that geometric spaces with constant non-zero curvature improve representation performance when the underlying graph structures of the data follow certain patterns \cite{liu2019hyperbolic,chami2019hyperbolic,gu2018learning}. For example, the hyperbolic space has a natural expressive ability for hierarchical structures, and the spherical space benefits for modeling cyclic structures.

\begin{figure}
  \centering
  \includegraphics[width=0.8\linewidth]{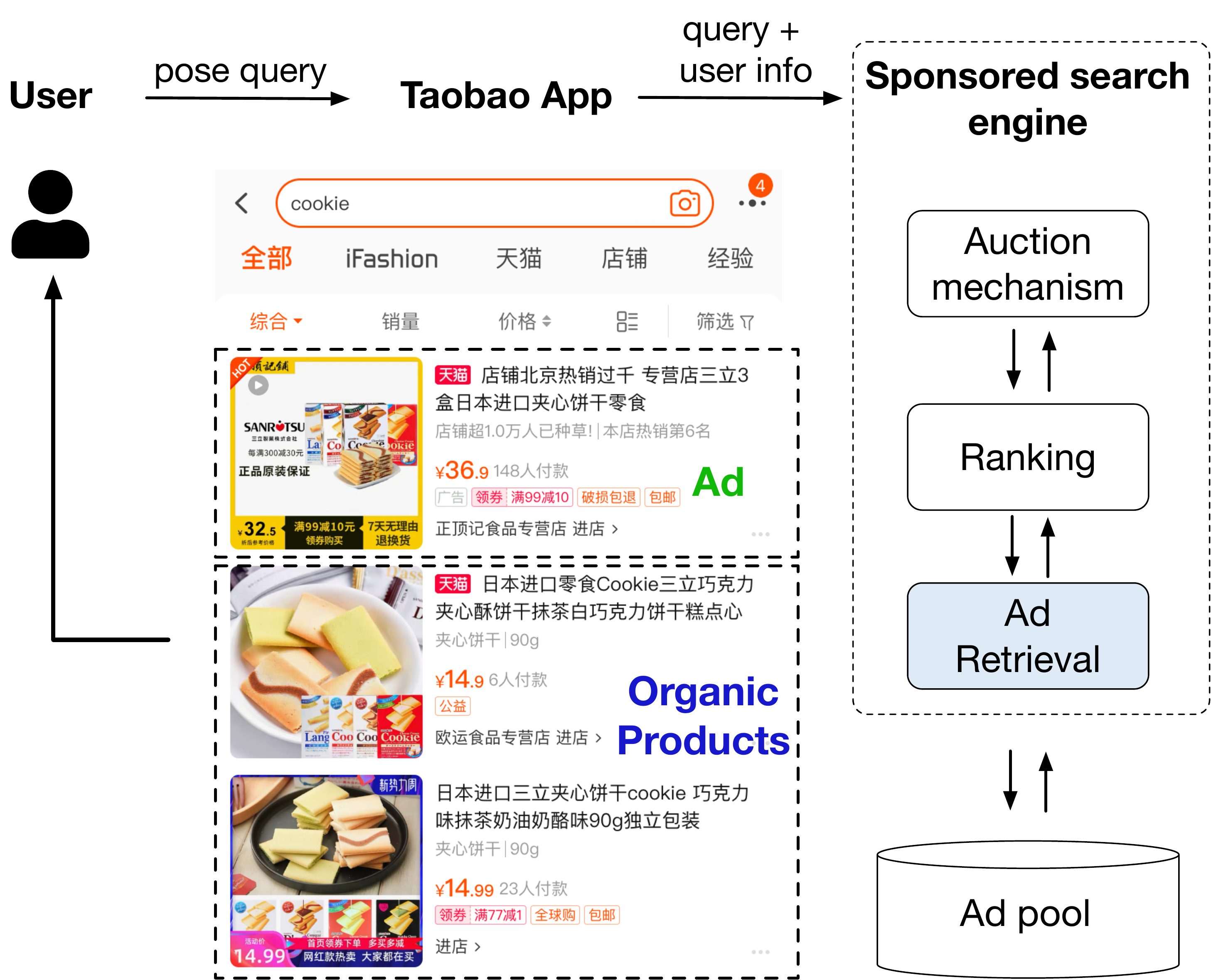}
  \caption{Illustration of the workflow of Taobao's sponsored search platform. }
  \label{fig:userplatforminter}
\end{figure}

However, there are still great challenges in non-Euclidean space representation learning on the complex heterogeneous graph:

\begin{itemize}
    \item \textit{A complex graph often has varied structures coupled together, and finding the optimal representation space is particularly challenging.} Traditional constant curvature space embedding methods only map nodes to a single space, so they suffer large distortion when representing varied structures. Existing mixed-curvature embedding methods \cite{gu2018learning,han2020dyernie,wang2021mixed} are all based on manually defined curved spaces, where inappropriate spaces will lead to failure to accurately capture structural information. However, space types and curvatures for complex graphs are difficult to predict, and it is also hard to manually find the optimal space mixture from the exponential combinations.
    \item \textit{Different types of nodes and edges on heterogeneous graphs often have distinct structural characteristics, and it is essential to express various types of relationships discriminatively.} In e-commerce scenario, queries often have hierarchical relationships, while items and ads tend to have stronger spherical characteristics. Existing approaches model different relationships between node pairs in the same mixed-curvature space, ignoring these heterogeneities. So they lack the ability to precisely express the structural characteristics.
\end{itemize}

With these challenges, in this paper, we present an Adaptive Mixed-Curvature ADvertisement retrieval system called \our{} for Taobao's sponsored search. The key novelty of \our{} lies in both the model design and system implementation. In terms of model design, we introduce a fully adaptive mixed-curvature embedding learning framework, which is composed of two adaptive components. i) A node-level mixed-curvature encoder maps graph entities with different local structures to the optimal mixed-curvature space. The space is composed of multiple subspaces of any type and curvature, which are jointly adapted in a data-driven manner. ii) An edge-level mixed-curvature scorer infers the linkage between two entities by projecting different types of relationships into a proper edge-wise mixed-curvature space, and then combining the linkage probability from the individual subspaces in an attentive manner. In terms of system implementation, we develop and deploy the ad retrieval system for both offline training and online serving. In offline experiments, \our{} outperforms existing geometric learning methods, which preliminarily proves the effectiveness of the system. For online serving, \our{} builds two-layer inverted indices to support efficient ad retrieval for tens of thousands of search requests per second. The online experiments prove that revenue has increased significantly in Taobao's sponsored search platform.

In summary, our main contributions are:
\begin{itemize}
    \item To the best of our knowledge, we are the first to apply mixed-curvature geometry to an industrial level system, including the offline model design and online retrieval system deployment.
    \item We propose an adaptive mixed-curvature representation learning model for complex heterogeneous graphs. A delicate node-level encoder maps different topological structures into an adaptive mixed-curvature space, which tackles the challenge of manually selecting the appropriate spaces. To dynamically distinguish heterogeneous edges, an edge-level scorer is designed in an attentive manner. 
    \item Extensive offline evaluations and online experiments are conducted to demonstrate the effectiveness of the proposed system. \our{} not only has guiding significance for future related work in the research field, but also brings great application value in industrial scenarios. 
\end{itemize}

%% file: problem.tex
\section{Problem Setup}

\subsection{Taobao Sponsored Search Platform}

As illustrated in Fig.~\ref{fig:userplatforminter}, in Taobao mobile app, a user can search products by posing queries to the system, and the system returns a set of products after retrieving posed queries. The results contain both organic products (items) and advertisements (ads). In practice, to balance between efficiency and effectiveness, the sponsored search platform generally employs a multi-stage search architecture. The sponsored search platform first retrieves a set of relevant ads from a large ad pool (\textit{ad retrieval}), then ranks the retrieved ads by elaborate models (\textit{ranking}) and compute the charging price to the advertiser by generalized second price auction mechanisms (\textit{auction mechanism}). In this paper, the proposed \our{} focuses on the \textbf{\textit{ad retrieval}} stage in the sponsored search platform.

\subsection{Problem Definition}

The user behavior logs in Taobao naturally forms a massive heterogeneous interaction graph $\mathcal{G} = (\mathcal{V}, \mathcal{E})$. $\mathcal{V}$ is the set of nodes, including queries $\mathcal{V}^q$, items $\mathcal{V}^i$ and ads $\mathcal{V}^a$, and $\mathcal{E}$ is the set of edges, whose relations include clicking, co-clicking, semantic and co-bidding. Given an online request consisting of a query $q$ posed by a user and a list of historical click items $\textbf{P} = (p_1, p_2, \ldots, p_m)$ under $q$, ad retrieval aims to get the most valuable ad sets $\{a\}$ with a matching function $f$
\begin{equation}
a = \underset{a \in \mathcal{V}^a}{\operatorname{argmax}} \ f(q, \textbf{P}, a|\mathcal{G})
\end{equation}
Hence, the system objective can be regarded as link prediction between $\mathcal{V}^q$, $\mathcal{V}^i$ and $\mathcal{V}^a$. Graph embedding has become one of the most advanced techniques in handling the link prediction task. Specifically, nodes $v \in \mathcal{V}$ are firstly projected as representations $\mathbf{v}$ with graph structures preserved, and then the matching function $f$ is learned as the distance between two node representations. Since some recent work has shown the superiority of non-Euclidean embedding on representing graph structures, in this paper, \our{} is further designed to represent nodes in a more suitable space to retrieve ads more accurately.

%% file: preliminary.tex
\section{Preliminary}

\begin{table}[]
    \centering
    \caption{Three different types of constant curvature spaces.}
    \label{tab:manifold_definition}
      \begin{tabular}{cccc}
        \toprule
        \textbf{Space}    & \textbf{Hyperbolic} & \textbf{Euclidean} &  \textbf{Spherical} \\ \midrule
        Notation  & $\mathbb{H}$ & $\mathbb{E}$ & $\mathbb{S}$ \\
        Curvature & $\kappa < 0$ & $\kappa = 0$ &  $\kappa > 0$ \\ 
        Visualization  &  \raisebox{-.5\height}{\includegraphics[height=.8cm]{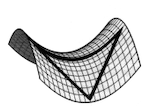}} & \raisebox{-.5\height}{\includegraphics[height=.8cm]{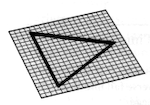}} & \raisebox{-.5\height}{\includegraphics[height=.8cm]{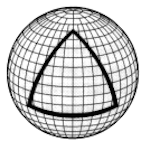}} \\  \bottomrule
      \end{tabular}
  \end{table}

In this section, we briefly introduce the background knowledge and notations related to \our{}. 
More details can be found in \cite{petersen2006riemannian}. 

\subsection{Constant Curvature Spaces} \label{sec:geometricspace}

Riemannian geometry, one of the most famous non-Euclidean geometries, has shown its higher capacity than traditional Euclidean geometry. Constant curvature space $\mathcal{M}^n_{\kappa}$ in Riemannian geometry, with dimension $n \geq 2$ and constant sectional curvature $\kappa \in \mathbb{R}$, has been widely studied due to its closed-form theory quantities. Table~\ref{tab:manifold_definition} intuitively illustrates three types of constant curvature spaces: hyperbolic $\mathbb{H}$ with $\kappa < 0$, Euclidean $\mathbb{E}$ with $\kappa = 0$ and spherical $\mathbb{S}$ with $\kappa > 0$. They provide benefits for modeling hierarchical \cite{sarkar2011low,candellero2016clustering,nickel2017poincare,sala2018representation}, grid \cite{monti2017geometric,bronstein2017geometric} and cyclical \cite{wilson2014spherical,liu2017sphere,xu2018shperical,gu2018learning,bachmann2020constant} data respectively. Though they have completely different properties, their stereographic projection models can be unified in the same model \cite{bachmann2020constant,kochurov2020hyperbolic}, which is called unified space $\mathbb{U}^n_{\kappa}$. Benefit from the fact that the curvature of $\mathbb{U}^n_{\kappa}$ varies from negative, zero and positive in a smooth way, the space can adaptively capture the hidden data structures without artificially determining the space type in advance. The basic operations for building non-Euclidean deep learning models in $\mathbb{U}^n_{\kappa}$ are summarized in Table \ref{tab:manifold_operation}, which are implemented via \textit{gyrovector spaces} framework \cite{ungar2008gyrovector, ganea2018hyperbolic} and different from traditional Euclidean operations.

\subsection{Mixed-curvature Space}

Since real industrial data shown in Fig.~\ref{fig:tree_cyclic_structure_sample} exhibits both tree and cyclical structures, space with multiple different curvatures, namely mixed-curvature space, is a better choice than constant curvature spaces. Product space proposed in \cite{gu2018learning} is a typical model of mixed-curvature space. Given a sequence of $N$ different constant curvature spaces $\mathcal{M}^{(1)}$, $\mathcal{M}^{(2)}$, ..., $\mathcal{M}^{(N)}$, the product space can be formalized as:
\begin{equation}
    \times_{i=1}^N \mathcal{M}^{(i)} = \mathcal{M}^{(1)} \times \mathcal{M}^{(2)} \times \cdots \times \mathcal{M}^{(N)}
\end{equation}
where $\times$ denotes the Cartesian product. The distance between two points $\mathbf{x}$ and $\mathbf{y}$ is defined as:
\begin{equation}
    \textrm{dist}(\mathbf{x}, \mathbf{y}) = \sum_{i=1}^N \textrm{d}_{\mathcal{M}^{(i)}}(\mathbf{x}_{\mathcal{M}^{(i)}}, \mathbf{y}_{\mathcal{M}^{(i)}})
\end{equation}
where $\mathbf{x}_{\mathcal{M}^{(i)}}$ and $\mathbf{y}_{\mathcal{M}^{(i)}}$ are the corresponding vectors on subspace $\mathcal{M}^{(i)}$ respectively. This traditional mixed-curvature space mainly has two drawbacks: the first is the need to manually select composed subspaces curvature, and the second is that the distance calculation simply sums up the distances without taking into account the importance of different subspaces. In this paper, we propose to represent nodes in adaptive mixed-curvature space, where each component is a unified space $\mathbb{U}^n_{\kappa}$ rather than a manually defined type, e.g., $\mathbb{H}$, $\mathbb{E}$ or $\mathbb{S}$. Moreover, each component is assigned with a learned weight, indicating the importance of each subspace in modeling graph structures. Thus \our{} is able to represent nodes in the best possible space to preserve complex structures in the interaction graph.

\begin{table*}[]
\renewcommand{\arraystretch}{1.5}
\small
\setlength{\tabcolsep}{2em}
\centering
 \caption{Summary of operations used in unified model $\mathbb{U}_\kappa^n$.}
 \label{tab:manifold_operation}
 \begin{tabular}{ccc}
    \toprule
    \textbf{Operation} & \textbf{Symbol} & \textbf{Closed-form Expression} \\ \midrule
    Exponential Map & $\exp_{\mathbf{x}}^{\kappa}(\cdot)$ & $\exp_{\mathbf{x}}^{\kappa}(\mathbf{v})={\mathbf{x}} \oplus_{\kappa}\left(\operatorname{\tan_\kappa}\left(\ \frac{\lambda_{\mathbf{x}}^{\kappa}\|{\mathbf{v}}\|_{2}}{2}\right) \frac{\mathbf{v}}{\|{\mathbf{v}}\|_{2}}\right)$\\ \hline
    Logarithmic Map & $\log_{\mathbf{x}}^{\kappa}(\cdot)$ & $\log_{\mathbf{x}}^{\kappa}(\mathbf{y})=\frac{2}{\lambda_{\mathbf{x}}^{\kappa}} \tan_\kappa^{-1}\left(\left\|-{\mathbf{x}} \oplus_{\kappa} {\mathbf{y}}\right\|_{2}\right) \frac{{-\mathbf{x}} \oplus_{\kappa} {\mathbf{y}}}{\left\|{-\mathbf{x}} \oplus_{\kappa} {\mathbf{y}}\right\|_{2}}$ \\ \hline
    Addition & $\oplus_{\kappa}$ & $\mathbf{x} \oplus_{\kappa} \mathbf{y}=\frac{\left(1-2\kappa\langle \mathbf{x}, \mathbf{y}\rangle-\kappa\|\mathbf{y}\|_{2}^{2}\right) \mathbf{x}+\left(1+\kappa\|\mathbf{x}\|_{2}^{2}\right) \mathbf{y}}{1-2 \kappa\langle \mathbf{x}, \mathbf{y}\rangle+\kappa^{2}\|\mathbf{x}\|_{2}^{2}\|\mathbf{y}\|_{2}^{2}}$ \\ \hline
    Matrix Multiplication & $\otimes_{\kappa}$ & $\mathbf{M} \otimes_{\kappa} \mathbf{x}=\exp_{\mathbf{0}}^\kappa \left( \mathbf{M} \log_{\mathbf{0}}^\kappa \left( \mathbf{x} \right) \right)$  \\ \hline
    Activation Function & $\sigma^{\kappa_1 \rightarrow \kappa_2}$ & $\sigma^{\kappa_1 \rightarrow \kappa_2} \left( \mathbf{x} \right )=\exp_{\mathbf{0}}^{\kappa_2} \left( \sigma \left(  \log_{\mathbf{0}}^{\kappa_1} \left( \mathbf{x} \right) \right) \right)$  \\ \hline
    Geodesics Distance & $\textrm{d}_\kappa(\cdot, \cdot)$ & $\textrm{d}_\kappa(\mathbf{x}, \mathbf{y})=2\operatorname{\tan}_\kappa^{-1}\left(\left\|-{\mathbf{x}} \oplus_{\kappa} {\mathbf{y}}\right\|_{2}\right)$ \\ \hline
    \multirow{4}{*}{\begin{tabular}[c]{@{}c@{}}Trigonometric\end{tabular}} & $\tan_\kappa^{-1}(\cdot)$ & $\tan_\kappa^{-1}(\mathbf{x})=\left\{\begin{array}{ll} \frac {1} {\sqrt {-\kappa}} \tanh ^{-1}\left(\sqrt {-\kappa} \mathbf{x} \right) & \kappa < 0 \\ \mathbf{x}-\frac{1}{3} |\kappa| \mathbf{x}^{3} & \kappa=0 \\ \frac {1} {\sqrt {\kappa}} \tan ^{-1}\left(\sqrt {\kappa} \mathbf{x} \right) & \kappa > 0 \end{array}\right.$ \\ & $\tan_\kappa(\cdot)$ & $\tan_\kappa(\mathbf{x})=\left\{\begin{array}{ll} \frac{1}{\sqrt{-\kappa}} \tanh \left(\sqrt{-\kappa} \mathbf{x} \right) & \  \ \ \  \kappa<0 \\ \mathbf{x} + \frac{1}{3} |\kappa| \mathbf{x}^{3} & \ \ \ \ \kappa=0 \\ \frac{1}{\sqrt \kappa} \tan \left(\sqrt{\kappa}\mathbf{x} \right) & \ \ \ \ \kappa > 0 \end{array}\right.$\\
    \bottomrule
 \end{tabular}
\end{table*}

%% file: model.tex
\section{\our{} Design} \label{sec:system}

\begin{figure}
  \centering
  \includegraphics[width=\linewidth]{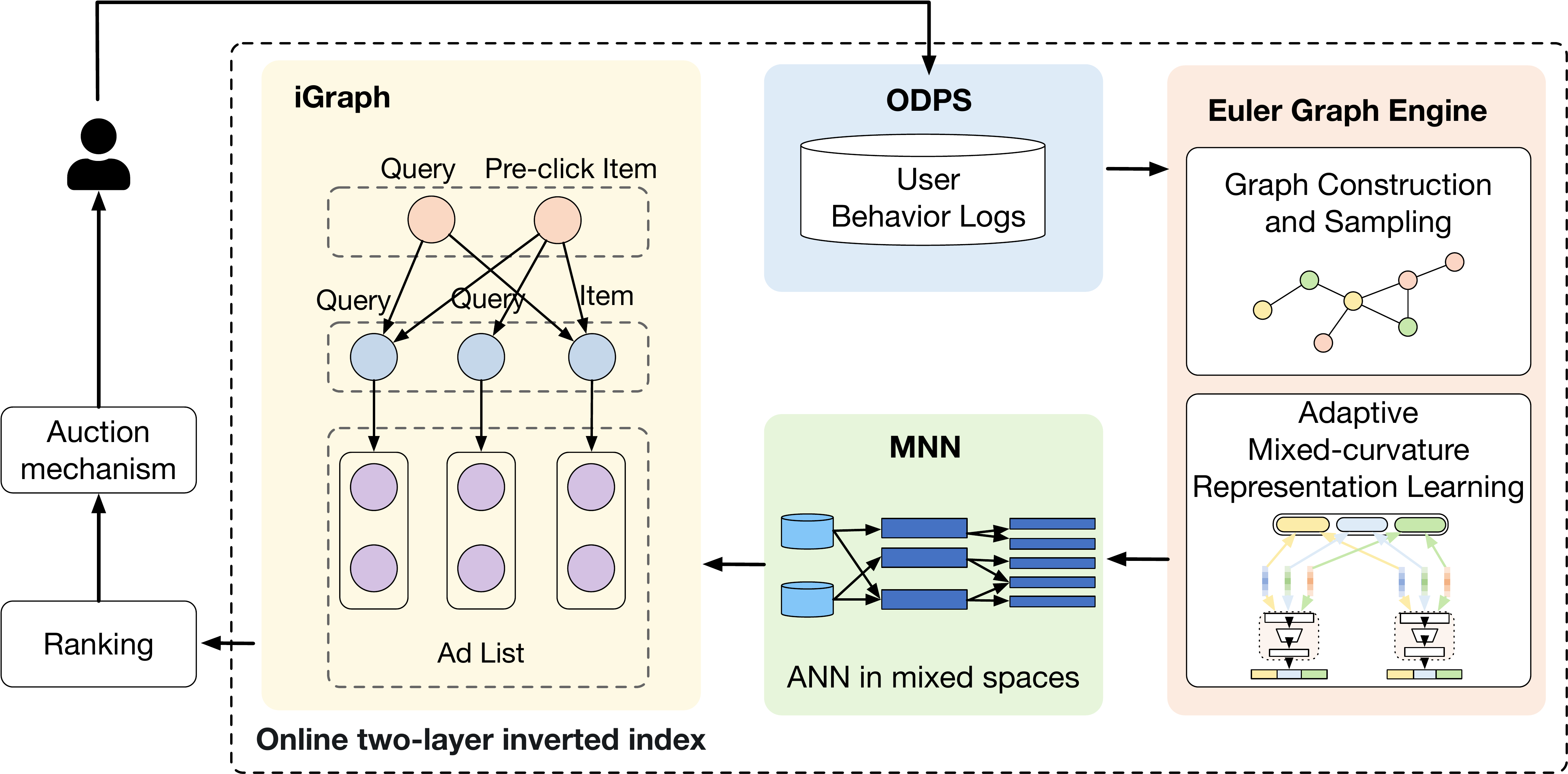}
  \caption{The overview of \our{} including model training, index building and online serving framework.}
  \label{fig:onlinesystem}
\end{figure}

The overview of \our{} is shown in Fig.~\ref{fig:onlinesystem}. The user behavior logs are stored in ODPS\footnote{ODPS is a general purpose, fully managed, multi-tenancy data processing platform for large-scale data warehousing developed by Alibaba. \url{https://www.alibabacloud.com/product/maxcompute}} (Open Data Processing Service) system. Then a large-scale heterogeneous graph is constructed based on these logs and imported into a distributed graph learning engine \textit{Euler}\footnote{Euler is a large-scale distributed graph learning framework developed by Alibaba with rich built-in state-of-the-art algorithms. \url{https://github.com/alibaba/euler}}. We also train our adaptive mixed-curvature model on this graph using Euler. After that the trained embeddings are fed to an efficient Mixed-curvature approximate Nearest Neighbor search module (\textit{MNN}) to generate the inverted index for online serving. When a user searches a query on Taobao app, the two-layer online ad retrieval module retrieves ads by searching over the inverted index in the \textit{iGraph}\footnote{iGraph is a large graph retrieval engine that supports online data updating and retrieval developed by Alibaba.} system.

In this section, we present the detailed design of \our{}. We first describe the process of constructing a heterogeneous graph and generating meta-path based training samples. Then we introduce the adaptive mixed-curvature learning approach, including the node-level encoder and edge-level scorer. Finally, we develop a fast \textit{ANN} search technique to speed up building the inverted indices, which are applied to our two-layer online ad retrieval system.

\subsection{Graph Construction and Sampling} \label{sec:graphconstruction}

\begin{figure*}
  \centering
  \includegraphics[width=0.85\linewidth]{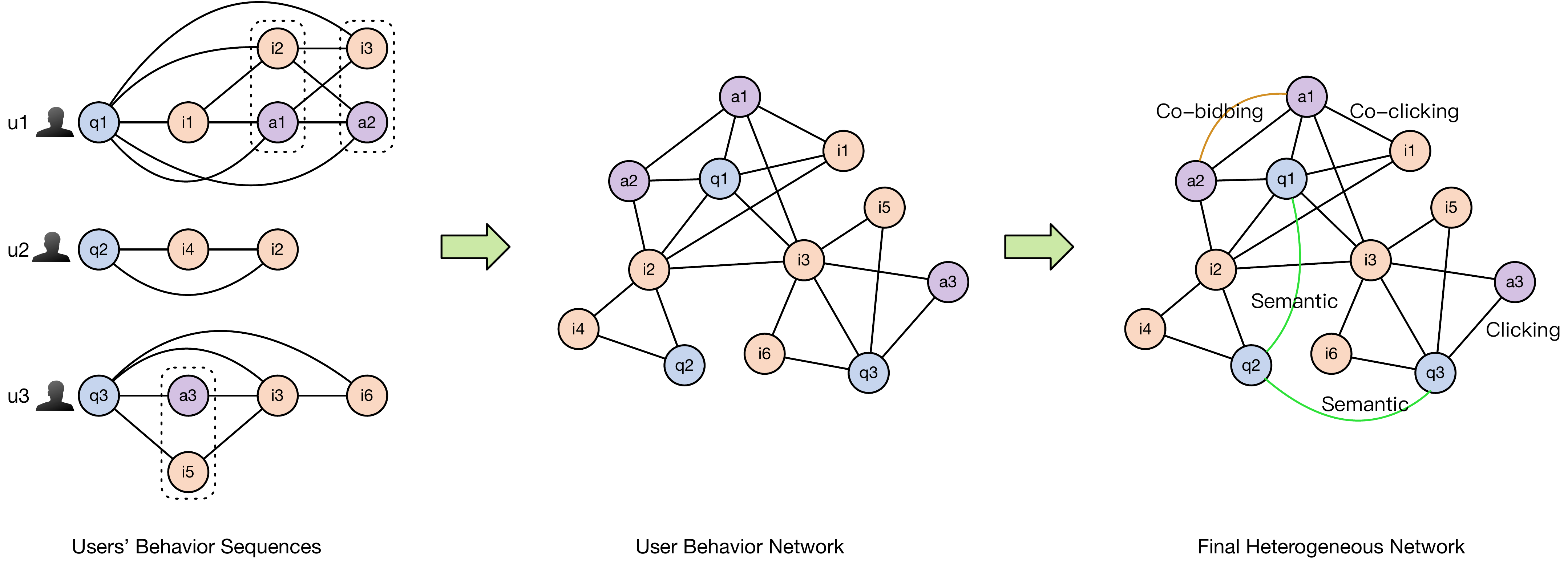}
  \caption{The pipeline of constructing the large-scale heterogeneous graph from massive user behaviors on Taobao's sponsored search platform.} 
  \label{fig:graph_construction}
\end{figure*}

\subsubsection{Graph construction} 

We elaborate the procedure to construct a heterogeneous network $\mathcal{G}$ from massive user behavior logs. As Fig.~\ref{fig:graph_construction} shows, we first construct the user behavior graph based on a large number of user search and click behavior sequences, and then add some other non-behavioral edges. Behavioral edges, i.e., clicking and co-clicking edges, can help to model popular nodes searched or clicked by many users, while non-behavioral edges, i.e., co-bidding edges and semantic similarity edges, provide more informative relations for cold nodes with sparse behavior links.

Concretely, the building process of these edges are depicted as follows: 

\begin{itemize}
  \item \textbf{Clicking/Co-clicking edges.} Taking Fig.~\ref{fig:graph_construction} as an example, a user $u_1$ searched query $q_1$ and then clicked item $i_1$, ad $a_1$ and $a_2$, where $a_1$ and $i_2$, $a_2$ and $i_3$ belong to the same product and are placed in the same dashed box. In this case, $q_1$ connects with $i_1$, $a_1$, $i_2$, $a_2$ and $i_3$ with clicking edges, and the adjacent items (or ads) are also connected with each other with co-clicking edges. Formally, given a clicking sequence $s = (s_1,..., s_m)$ under a user's search query $q$, we generate a clicking edge between each clicked node $s_i$ and the query node $q$, and a co-clicking edge between two adjacent clicked item (or ad) nodes $s_i$ and $s_j$.
  \item \textbf{Semantic similarity edges.} This type of edge links two queries together when their Jaccard similarity measured by textual terms is larger than a given threshold. By the aid of the textual contents, new nodes introduced to the system can get a better cold start.
  \item \textbf{Co-bidding edges.} In the sponsored search scenario, the advertisers usually specify a set of $<keyword, price>$ pairs for each ad to express the search traffic they are willing to reach and the amount of money they want to bid when the ads are clicked by users. Two ads are linked together if they are bidden with at least one same keyword.
\end{itemize}

\subsubsection{Meta-path based sample generation} 

In order to capture the high-order heterogeneous relations between different types of nodes, we adopt a meta-path sampling \cite{dong2017metapath2vec} approach to sample positive training data for the model. We first perform a random walk strategy along the defined meta-path on the heterogeneous graph, and based on the walked node sequences, the positive sample pairs are extracted by a sliding window. Table~\ref{tab:positive_sample_generate} illustrates the meta-path definitions to generate different positive node pairs. Note that in e-commerce platforms, the products are generally arranged in a category tree \cite{gao2020deep} with each product belonging to one leaf node in the category tree. To guarantee the relevance, we also require the sampled positive node pairs to be in the same category. The category of an item or ad is specified by the owner in advance. The category of a query is predicted by a sophisticated developed classification model. Given a positive node pair, we sample both hard negative and easy negative nodes through random sampling. Specifically, hard nodes are sampled from the same category as the positive ones while the easy nodes are sampled under the other categories. By using the hard negative samples, the model tends to be more robust and the learned representation can distinguish nodes at a finer granularity.

\begin{figure*}
  \centering
  \includegraphics[width=\linewidth]{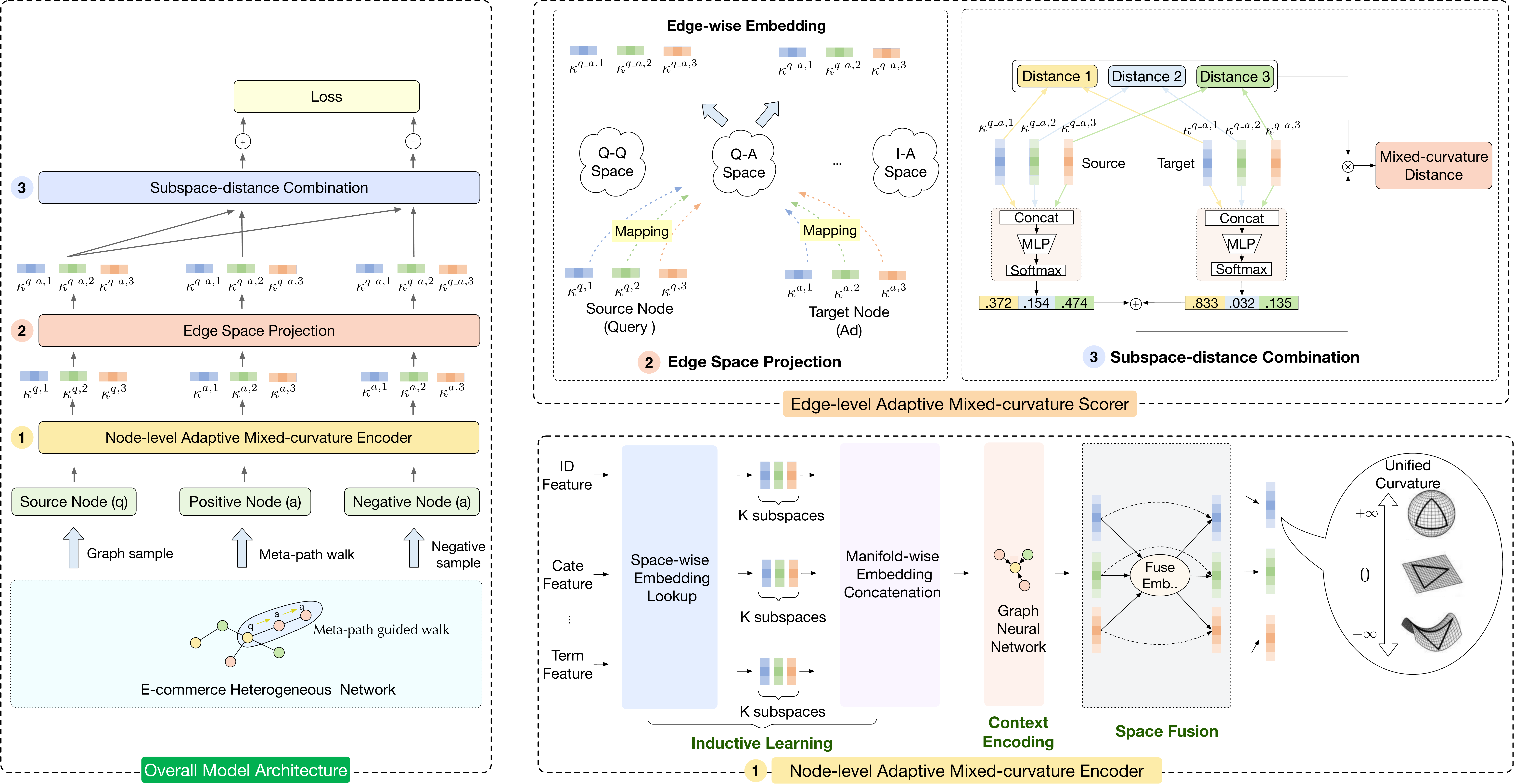}
  \caption{The architecture of the adaptive mixed-curvature representation model. The left part is the overall architecture and the right part depicts the main components including (1) node-level adaptive mixed-curvature encoder, (2) edge space projection and (3) subspace-distance combination. Note that (2) and (3) are merged into edge-level adaptive mixed-curvature scorer in the main text.} 
  \label{fig:overall_model}
\end{figure*}

\begin{table}
 \setlength{\tabcolsep}{0.45em}
 \footnotesize
  \caption{The generation process of positive node pairs using meta-path guided random walk.} 
  \label{tab:positive_sample_generate}
  \begin{tabular}{ccc}
    \toprule
    \textbf{Meta-path} & \textbf{Node Sequence} & \textbf{Positive Node Pairs}\\
    \midrule
    $q \xrightarrow{co-click} q \xrightarrow{semantic} q$ & $q_1 \rightarrow q_2 \rightarrow q_3$ & $<q_1, q_2>, <q_1, q_3>$ \\
    \hline
    $q \xrightarrow{click} i \xrightarrow{co-click} i$ & $q_1 \rightarrow i_1 \rightarrow i_2$ & $<q_1, i_1>, <q_1, i_2>$ \\
    \hline
    $q \xrightarrow{click} a \xrightarrow{co-bid} a$ & $q_1 \rightarrow a_1 \rightarrow a_2$ & $<q_1, a_1>, <q_1, a_2>$ \\
    \hline
    $i \xrightarrow{click} q \xrightarrow{semantic} q$ & $i_1 \rightarrow q_1 \rightarrow q_2$ & $<i_1, q_1>, <i_1, q_2>$ \\
    \hline
    $i \xrightarrow{co-click} i \xrightarrow{co-click} i$ & $i_1 \rightarrow i_2 \rightarrow i_3$ & $<i_1, i_2>, <i_1, i_3>$ \\
    \hline
    $i \xrightarrow{co-click} a \xrightarrow{co-bid} a$ & $i_1 \rightarrow a_1 \rightarrow a_2$ & $<i_1, a_1>, <i_1, a_2>$ \\
  \bottomrule
\end{tabular}
\end{table}

\subsection{Adaptive Mixed-curvature Representation Model} \label{sec:modelarchitecture}

In this section, we present the detailed model design of the proposed \our{}. As shown in Fig.~\ref{fig:overall_model}, it has two key components: a node-level adaptive mixed-curvature encoder and an edge-level adaptive mixed-curvature scorer. The former generates node representations in the adaptive mixed-curvature space to automatically learn useful information from a variety of complex structures, and the latter calculates similarities between heterogeneous nodes to model the different relations in the mixed-curvature space accurately.

\subsubsection{\textbf{Node-level adaptive mixed-curvature encoder}} \label{sec:encoder}

In order to extract useful information from a variety of complex structures, we encode heterogeneous entities into the adaptive mixed-curvature spaces, where optimal subspaces types and curvatures can be learned automatically. Specifically, we define the adaptive mixed-curvature space as the product of multiple unified manifolds $\mathbb{U}$, such that space can be learned to be either hyperbolic, spherical or Euclidean in a data-driven manner without additional data analysis and prior knowledge. As shown in Fig.~\ref{fig:overall_model}, the node-adaptive mixed-curvature encoder consists of the following stages: i) \textit{Inductive learning} maps the node features into an initial adaptive mixed-curvature embedding; ii) \textit{Context encoding} utilizes GCN \cite{ying2018graph} to make full use of the context information; iii) \textit{Space fusion} boosts the representations by aggregating embeddings from different subspaces.

\begin{table}[]
\centering
\caption{Node Types and Feature Types.} \small
\label{tab:node_type_and_feature}
\begin{tabular}{|c|m{6cm}<{\centering}|}
    \hline
    \textbf{Node} & \textbf{Feature Types} \\
    \hline
    Query & ID, Category, Terms\\
    \hline
    Item & ID, Category, Title terms, Brand, Shop \\
    \hline
    Ad & ID, Category, Title terms, Bidding words, Brand, Shop \\
    \hline
\end{tabular}
\end{table}

\paragraph{Inductive learning} To make the model inductive, we leverage the rich node features as the input to the node-level adaptive mixed-curvature encoder. Table~\ref{tab:node_type_and_feature} summarizes the features for different types of nodes. The embedding representation generated by the encoder lies in a product of $M (M \geq 1)$ unified curvature spaces, i.e., $\mathcal{U} = \times_{m=1}^M \mathbb{U}_m^{d}$, where $\times$ denotes the Cartesian product and $d$ denotes the dimension of each space. As shown in Fig.~\ref{fig:overall_model} and Eq.~\ref{eq:featureconv}, given a graph node $v_t \in \mathcal{V}$ with type $t \in T=\{query, item, ad\}$ and $k$ features $(f_1^{t}, f_2^{t}, \ldots, f_k^{t})$, we generate feature embeddings $\textbf{e}_j^{m,t}~(m=1, \ldots M)$ for each feature $f_j^{t}$ in each space. A product of $M$ Riemannian embeddings are gathered as the concatenation of feature embeddings through the exponential transformation in each space.  For the $m$-th geometric space with type $t$, we denote $\kappa^{m,t}$ as the space curvature and $\mathbf{x}^{m,t}$ as the Riemannian embedding in the space. Thus the embedding procedure is formulated as
\begin{equation}
\begin{aligned}
\label{eq:featureconv}
    v_t= \times_{m=1}^M \textbf{x}^{m,t} =  \times_{m=1}^M \exp^{\kappa^{m,t}}_{\boldsymbol{0}} \left( \left[ \textbf{e}^{m,t}_{1} || \ldots ||  \textbf{e}^{m,t}_{k} \right] \right)
\end{aligned}
\end{equation}
where $||$ denotes the concatenation operator.

\paragraph{Context encoding} To capture the context information, we utilize the GNN approach to aggregate the neighbor information. However, different from existing GNN methods \cite{hamilton2017inductive,velivckovic2018graph}, our node representation is a product of multiple embeddings, and moreover, different node types are represented in different sets of geometric spaces. Thus we project the neighbors' embeddings to the tangent space at the origin for each geometric space and graph convolution can be operated in a common space. Concretely, given a node $v_t$ of type $t$, let the representation of the first graph convolution layer $\mathbf{h}^{m,t,0} = \mathbf{x}^{m,t}$ and its neighbors with type $t'$ denoted by $\mathcal{N}^{t'}_{v}$. Then at layer $l>0$, the aggregated information $\mathbf{\hat{h}}^{m,t,l}$ after message passing is obtained as:
\begin{equation}
\begin{aligned}
\mathbf{\hat{h}}^{m,t,l} = 
& \sum_{t' \in T} \frac {1} {|\mathcal{N}^{t'}|} \sum_{j \in \mathcal{N}^{t'}}  
\log^{\kappa^{m,t'}}_{\boldsymbol{0}}\left(\mathbf{h}^{m,t',l-1}_{j}\right) \\
& \big|\big| \log^{\kappa^{m,t}}_{\boldsymbol{0}}\left(\mathbf{h}^{m,t,l-1}\right)
\end{aligned}
\end{equation}
Then updating the representation in $m$-th subspace with aggregated neighbor information is realized by a exponential map and a non-linear transformation. Formally, the node representation $\mathbf{h}^{m,t,l}$ at layer $l>0$ is updated as:
\begin{equation}
\mathbf{h}^{m,t,l} = 
\sigma^{\kappa^{m,t} \rightarrow \kappa^{m,t}}
\left(
\mathbf{W}^{m,t,l} \otimes_{\kappa^{m,t}}
\exp^{\kappa^{m,t}}_{\boldsymbol{0}} 
\left(
\mathbf{\hat{h}}^{m,t,l}
\right)
\right)
\end{equation}
The full representation of node $v_t$ after $L$ round graph convolution is denoted by  $v_{t} = \times_{m=1}^M \textbf{x}^{m,t} = \times_{m=1}^M \textbf{h}^{m,t,L}$.

\paragraph{Space fusion} To enhance the learning of each subspace and capture the global structural information more effectively, we propose a fusion mechanism by aggregating subspaces representations in the global tangent space. Formally, for the $m$-th subspace embedding $\textbf{x}^{m,t}$ of node $v_{t}$, the global fused representation $\mathbf{h}^{t}$ is obtained from the average pooling of subspaces embeddings:
\begin{equation}
    \mathbf{h}^{t} = \frac{1}{M} \sum_{m=1}^M \log_{\boldsymbol{0}}^{\kappa^{m,t}} \left( \textbf{x}^{m,t} \right)
\end{equation}
Then the interacted representation in each subspace is derived as:
\begin{equation}
    \textbf{x}^{m,t} = \exp^{\kappa^{m,t}}_{\boldsymbol{0}} \left(\mathbf{W}^{m,t}_{1} \cdot \left[ \textbf{h}^{t} || \log_{\boldsymbol{0}}^{\kappa^{m,t}}(\textbf{x}^{m,t}) \right] \right)
\end{equation}
where the information is linearly combined from both the current subspace embedding and the global fused information.

\subsubsection{\textbf{Edge-level adaptive mixed-curvature scorer}} \label{sec:distcalc}

Since each relation has its unique characteristics, to model the heterogeneity in the mixed-curvature space, we map different relations into different spaces, and the space weights between two nodes are calculated based on the attention mechanism. The edge-level adaptive mixed-curvature scorer in Fig.~\ref{fig:overall_model} estimates the probability of edge existence between two nodes by first mapping their node-level embeddings to edge-level adaptive mixed-curvature spaces (\textit{edge space projection}), and computes the probability by assembling each subspace distance (\textit{subspace-distance combination}). Specifically, given a source node $x_{t_1}$ of type $t_1$, whose representation generated by node-level encoder is $\times_{m=1}^M \textbf{x}^{m,t_1}$ and a target node $y_{t_2}$ of type $t_2$ with representation $\times_{m=1}^M \textbf{y}^{m,t_2}$, their link probability is predicted in the following steps:

\paragraph{Edge space projection} Since $\textbf{x}^{m,t_1}$ and $\textbf{y}^{m,t_2}$ are learned in different geometric spaces, geodesic distance cannot be derived directly. Thus we project them to a best common edge-level geometric space with curvature $\kappa^{m,r}$ indexed by the $r$-th edge type and the $m$-th subspace. Note that the same node would be mapped to different edge-level spaces due to heterogeneity. The mapping function is defined as:
\begin{equation}
\textrm{proj}_{r}(\textbf{x}^{m,t}) = 
\sigma^{\kappa^{m,t} \rightarrow \kappa^{m,r}}
\left(
\mathbf{W}^{m,t}_{2} \otimes_{\kappa^{m,t}}
\textbf{x}^{m,t}
\right)
\end{equation}
where $\mathbf{W}^{m,t}_{2}$ denotes the linear transformation matrix. Then the distance between subspace embedding $\textbf{x}^{m,t_1}$ and $\textbf{y}^{m,t_2}$ is computed as:
\begin{equation}
    \textrm{d}_m(\textbf{x}^{m,t_1}, \textbf{y}^{m,t_2}) = \textrm{d}_{\kappa^{m,r}} \left( \textrm{proj}_{r}(\textbf{x}^{m,t_1}), \textrm{proj}_{r}(\textbf{y}^{m,t_2}) \right)
\end{equation}
where $\textrm{d}_{\kappa^{m,r}}$ denotes the geodesics distance in the space with curvature ${\kappa^{m,r}}$ defined in Table \ref{tab:manifold_operation}.

\paragraph{Subspace-distance combination} Noticing that the complex structures are not uniformly distributed among the whole graph, we use an attention-based mechanism to combine subspaces with dynamic weights. Different from \cite{wang2021mixed} which used fixed global space weights for all pairs of nodes, the subspace weights in our approach are learned in a dynamic way and vary with different node pairs. As illustrated in Fig.~\ref{fig:overall_model}, the $m$-th subspace weight between two nodes $x_{t_1}$ and $y_{t_2}$ is calculated as:
\begin{equation}
 \textrm{w}(\textbf{x}^{m,t_1}, \textbf{y}^{m,t_2}) = \textrm{w}'(\textbf{x}^{m,t_1}) + \textrm{w}'(\textbf{y}^{m,t_2})
\end{equation}
where $\textrm{w}'(\textbf{x}^{m,t_1})$ and $\textrm{w}'(\textbf{y}^{m,t_2})$ are the node level subspace attention weights and they are summed up to get the global subspace weight for the node pair. Thus the weights can be pre-calculated before performing \textit{MNN} retrieval. Take node type $t$ as an example, $\textrm{w}'(\textbf{x}^{m,t})$ is formulated as:
\begin{equation}
     \boldsymbol{\alpha}^t = \mathbf{W}^{t} \cdot \left[  \textrm{proj}_{r}(\textbf{x}^{1,t}) || \ldots || \textrm{proj}_{r}(\textbf{x}^{M,t}) \right]
\end{equation}
\begin{equation}
\textrm{w}'(\textbf{x}^{m,t}) = \textrm{softmax} \left( \boldsymbol{\alpha}^t \right)=\frac{\textrm{exp}\left( \boldsymbol{\alpha}^t_m \right)}{\sum_{i=1}^M \textrm{exp}\left(\boldsymbol{\alpha}^t_i\right)}
\end{equation}
in which $\boldsymbol{\alpha}^t$ is a matrix of size $1 \times M$ and each element represents the importance of the corresponding subspace. The final distance between $x_{t_1}$ and $y_{t_2}$ is obtained by multiplying the weights and distances of each subspace:
\begin{equation}
    \textrm{dist}(x_{t_1}, y_{t_2}) = \sum_{m=1}^M \textrm{w}(\textbf{x}^{m,t_1}, \textbf{y}^{m,t_2}) \cdot \textrm{d}_m(\textbf{x}^{m,t_1}, \textbf{y}^{m,t_2})
\end{equation}

\subsubsection{\textbf{Loss}} \label{sec:training}

A training sample $<x_{src}, x_{pos}, \{ x_{neg_i} \}_{i=1}^K>$ consists of a source node, a positive node and $K$ sampled negative nodes. All training samples of different relation types are learned jointly. Standard triplet loss is used as:
\begin{equation}
    \mathcal{L} = \frac{1}{K} \sum_{i=1}^K \left[ m + \operatorname{sim}\left(x_{src}, x_{pos} \right ) - \operatorname{sim}\left(x_{src}, x_{neg_i} \right )  \right ]_+
\end{equation}
where $\left[ a \right ]_+=\max \left(0,a \right )$ is the hinge function. $m>0$ is the margin value and is set to $0.5$ empirically. The $\operatorname{sim}$ function transforms the node distance to link probability via the Fermi-Dirac distribution \cite{nickel2017poincare} defined as $\operatorname{sim}\left(x, y \right ) = \sigma(t(r-\textrm{dist}(x, y)))$, in which $\sigma$ is the sigmoid function and $r, t$ denotes the radius and temperature factor respectively. In practice $r$ is set to $1$ and $t$ is set to $5$ to get the best performance. 

 \begin{figure}
  \centering
  \includegraphics[width=\linewidth]{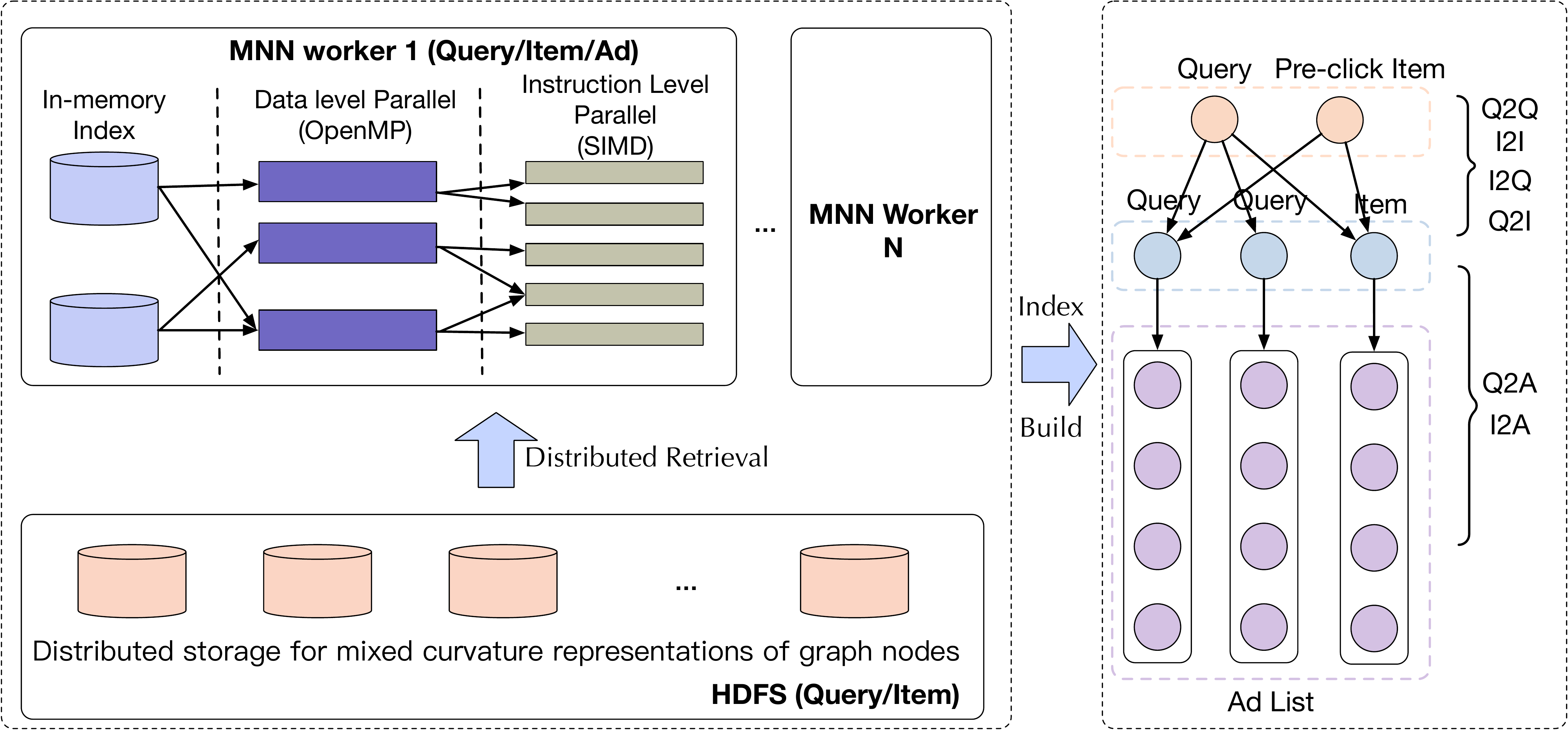}
  \caption{Illustration of \textit{MNN} and the two-layer inverted index used for online ad retrieval.}
  \label{fig:impoptimization}
\end{figure}

\subsection{Two-layer Online Retrieval} \label{sec:online}

Given an online request consisting of a query $q$ posed by a user $u$ and a list of historical click items $\textbf{P} = (p_1, p_2, \ldots, p_m)$ under $q$, the ad retrieval system aims to search a set of ads relevant to $q$ and preferred by $u$ efficiently. To achieve this goal, we first generate six types of inverted indices using the trained graph node representations. Then we deploy a two-layer online ad retrieval framework based on the constructed indices to receive user search requests and efficiently retrieve relevant ads.

\subsubsection{Offline inverted index construction via MNN}

As introduced in the previous section, we require to build six different kinds of inverted indices (i.e., Q2Q, Q2I, I2Q, I2I, Q2A and I2A) to conduct online ad retrieval. In each index, the key is the graph node and the value is the retrieved nearest $K$ corresponding results. Taken Q2Q as an example, the inverted indices of query nodes $x_q$ is calculated as $\operatorname*{argmin}_{y_q \in \mathcal{V}^q} \operatorname{dist}(x_q,y_q)$. Different from traditional metrics such as dot product or cosine in Euclidean space, the similarity between two nodes in our approach is calculated based on the attention mechanism, which is more complex and hard to directly use traditional nearest neighbor search approach such as product quantification \cite{jegou2010product}. To speed up the index construction for our model, as shown in Fig.~\ref{fig:impoptimization}, we develop a two-level parallelism architecture by distributing mixed-curvature representations of graph nodes to a set of workers (GPU machines). The embeddings of the keys are distributed stored in HDFS\footnote{A distributed file system designed to run on commodity hardware. \href{https://hadoop.apache.org/docs/current/hadoop-project-dist/hadoop-hdfs/HdfsDesign.html}{https://hadoop.apache.org/docs/current/hadoop-project-dist/hadoop-hdfs/HdfsDesign.html}} files and each worker stores the embeddings of all the nodes. Moreover, the parallelism of the nearest neighbor computation is carried out at both data level (OpenMP \cite{chandra2001parallel}) and instruction level (SIMD \cite{zhou2002implementing}) inside each worker. After these effective optimizations, the inverted indices can be built in less than two hours for 100 million nodes, which is acceptable in industrial scenarios.

\subsubsection{Online two-layer ad retrieval}

Due to the latency and computational cost constraint of online ad retrieval, directly performing model inference online is infeasible. To address this issue, the generated inverted indices above are utilized to build a two-layer ad retrieval system as shown in Fig.~\ref{fig:impoptimization}. The first layer expands the raw search query $q$ and pre-click items $\textbf{P}$ to a set of highly related queries and items through Q2Q, Q2I, I2Q and I2I indices, and the second layer aims to retrieve a collection of related ads by the keys through Q2A and I2A indices. Different from traditional embedding based methods such as \cite{huang2020embedding} which use queries to retrieve ads directly, our two-layer retrieval architecture can make use of more user input signals and rewrite to more effective keys, allowing the system to cover more online traffic.

%% file: implementation.tex
\section{\our{} Deployment}

To deploy \our{}, distributed storage and deep learning framework are utilized to handle the massive interaction graph data. Several techniques are developed to stabilize the training process. Meanwhile, incremental learning is introduced to save training time and resources.

\subsection{Distributed storage and training} The massive graph data is distributed stored in the Euler engine. Specifically, graph nodes are scattered in several shards by applying a hash function to their IDs. Nodes with the same hash value are stored in the same shards, as well as their attributes and neighborhoods. We use the alias method \cite{walker1977efficient} to conduct effective negative sampling with constant time complexity. To handle high-dimensional sparse data, we adopt the open-source deep learning framework XDL \cite{jiang2019xdl} as the training engine. \our{} uses the parameter server paradigm which can store the parameters and embeddings in a distributed manner. In each training iteration, the XDL worker first requests the Euler engine to generate samples through meta-path based walking and negative sampling, then pushes the updated parameters to the parameter server in asynchronized mode to achieve higher system throughput.

\subsection{Training stability} \label{sec:training_stability} In industrial scenarios with tremendous data, numerical instability is ubiquitous especially when training curved models. The reasons are mainly two-fold: i) Due to the limited radius of hyperbolic spaces, the embeddings are likely pushed out of the valid region by gradients descent operations; ii) Since mapping functions $\exp_{\mathbf{x}}^{\kappa}(\cdot)$ and $\log_{\mathbf{x}}^{\kappa}(\cdot)$ are commonly used, parameters lying on the steep zone of these functions may suffer from gradient exploding and vanishing problems. To tackle the out of boundary issue, we introduce a curved space regularization term to enable the nodes to be distributed around numerical stable zones:
\begin{equation}
    \mathcal{L}_{reg} = \sum\nolimits_{x \in <x_{src}, x_{pos}, \{ x_{neg_i} \}_{i=1}^K>}\textrm{dist}(x,\boldsymbol{0})
\end{equation}
aiming to minimize the distance between each node and the origin. The regularization weight is set to 0.001 empirically. For the gradient exploding and vanishing problem, we use the gradient clipping technique to avoid abnormal gradients and apply the learning rate warm-up strategy to stabilize the parameters update.

\subsection{Incremental learning} In a large e-commerce platform like Taobao, billions of graph nodes and edges are generated from user behavior logs every day. Batch training using the data within a time window (e.g., 7 days) is time-consuming. To tackle this challenge, we introduce the day-level incremental training for acceleration. Specifically, the model first inherits the existing previous day's model and is further trained based on the new logs. Since the features follow the long-tailed distribution, a large proportion of the features occur rarely but occupy a large amount of space. To avoid model size expansion, we design an LRU (least recently used) based feature exit mechanism, to eliminate the features that do not appear for a period of time. Benefiting from the massive amount of data, the metrics of \our{} are relatively smooth every day even during certain special days (e.g., Chinese Spring Festival).

%% file: experiments.tex
\section{EXPERIMENTS AND RESULTS}

In this section, we evaluate both the offline model and the whole online system. Comparison and ablation studies are conducted to answer the following questions:
\begin{itemize}
    \item \textbf{RQ1.} How does the model of our proposed \our{} perform compared to the recently proposed methods?
    \item \textbf{RQ2.} Where does the performance gain come from?
    \item \textbf{RQ3.} How do the key hyper-parameters impact \our{} performance?
    \item \textbf{RQ4.} Is the system efficient and scalable for large-scale data?
    \item \textbf{RQ5.} How does the whole system perform in real-time billions of e-commercial search traffic?
\end{itemize}

\subsection{Experimental Setup}

\subsubsection{Dataset}
We collect user behavior logs on Taobao's sponsored search platform to construct the heterogeneous graph for evaluation. The graph consists of three types of nodes, i.e., query, item and ad, and three types of edges, i.e., clicking/co-clicking edge, semantic similarity edge and co-bidding edge. The details of graph construction are shown in section \ref{sec:graphconstruction}. Note that we use user behavior logs in one day for offline evaluation and logs in seven days for the online A/B experiment. The statistics of graph nodes and edges collected from different days are shown in Table~\ref{tab:dataset_statistic}.

\begin{table}[]
\setlength{\tabcolsep}{0.9em}
\centering
\caption{Statistics of Taobao's sponsored search data. Logs in 1 day are used for offline evaluation. Logs in 7 days are used for online serving.} 
\small
\label{tab:dataset_statistic}
\begin{tabular}{cP{1.1cm}P{1.1cm}P{1.1cm}c}
\toprule
\textbf{Logs} & \textbf{\#Nodes (Query)} & \textbf{\#Nodes (Item)} & \textbf{\#Nodes (Ad)} & \textbf{\#Edges} \\ \midrule
1 day & 40M & 60M & 6M & 5.3B \\
7 days & 150M & 140M & 10M & 30.8B \\
\bottomrule
\end{tabular}
\end{table}

\subsubsection{Baselines}

We compare our method \our{} with the following state-of-the-art methods:
\begin{itemize}
    \item Euclidean models including homogeneous and heterogeneous methods, \textit{i.e.}, DeepWalk \cite{perozzi2014deepwalk}, LINE \cite{tang2015line}, Node2Vec \cite{grover2016node2vec}, Metapath2Vec \cite{dong2017metapath2vec}. To quantify the gain from model space, we also implement \our{} in Euclidean space without changing the model architecture, denoted by $\rm{\our{}^\mathbb{E}}$.
    \item Constant curvature models, \textit{i.e.}, HyperML \cite{vinh2020hyperml}, HGCN \cite{chami2019hyperbolic}. For fair comparision, we also implement \our{} in hyperbolic, spherical and unified space, denoted by $\rm{\our{}^\mathbb{H}}$, $\rm{\our{}^\mathbb{S}}$ and $\rm{\our{}^\mathbb{U}}$ respectively.  
    \item Mixed-curvature models, \textit{i.e.}, GIL \cite{zhu2020graph}, $\rm{M^2GNN}$ \cite{wang2021mixed}, and product space \cite{gu2018learning}. Specifically, we try all possible manifold combinations for product space and report the best result as $\rm{Product^{best}}$.
\end{itemize}

\subsubsection{Implementation Details} 

Since model parameters are located in tangent space, we use vanilla AdaGrad optimizer \cite{duchi2010adaptive} to train the model. Curvature parameters in different model layers are trainable. Hyperparameters \{batch size, embedding dimension,  negative sample size\} are chosen empirically as \{1024, 120, 6\}. The ratio of easy negative samples to hard negative samples is set to 2:1. Grid searching is performed on hyperparameters \{submanifold number, learning rate\} to get the best configuration as \{2, 1e-2\}. We conduct the experiments on 200 workers and 20 parameter servers. Since embedding hashing consumes most of the training time,  a total of 2400 CPU cores are used without GPU usage. As mentioned in \cite{nickel2018learning}, numerical instabilities arise from poincar{\'e} ball distance especially in massive parallel optimization, strategies in \ref{sec:training_stability} are applied to stabilize the training process.

\subsubsection{Evaluation Metrics}

We train our model on the one day's graph and evaluate the performance on the next day's graph. We use Next AUC (area under the roc curve evaluated on the next day data), Hitrate@$K$ and nDCG@$K$ as the evaluation metrics. Since different approaches adopt different random walk strategies to generate samples, for comparability, we report the Next AUC results for the methods using the same samples with our method. For Hitrate and nDCG metrics, we use the item/ad list sorted by click number under each query as the ground truth. Due to the space limitation, we only report metrics on query-to-item (Q2I) and query-to-ad (Q2A) relations.

\subsection{RQ1: Offline Results}

\begin{table*}[]
\centering 
\footnotesize
\setlength{\tabcolsep}{0.35em}
\caption{Experimental results (Next AUC, Training Time, Hitrate@$K$ and nDCG@$K$) on graph data collected from Taobao's E-commerce platform. $\mathbb{E}$ represents the Euclidean models, $\mathbb{C}$ represents the constant curvature models and $\mathbb{M}$ denotes the mixed-curvature models. The best results are in bold. Relative improvement compared to Euclidean space without the changing of model architecture is shown in the last row.}
\label{tab:overall_results}
\begin{tabular}{cccccccccccccccccc}
\toprule
\multirow{3}{*}{\textbf{$\mathcal{M}$}} & \multirow{3}{*}{\textbf{Model}} & \multirow{3}{*}{\begin{tabular}[c]{@{}c@{}}\textbf{Next}\\ \textbf{AUC}\end{tabular}} &
\multirow{3}{*}{\begin{tabular}[c]{@{}c@{}}\textbf{Training}\\ \textbf{Time (h)}\end{tabular}} & & \multicolumn{6}{c}{\textbf{Q2I}} &  & \multicolumn{6}{c}{\textbf{Q2A}} \\ \cline{6-11} \cline{13-18} 
 &  &  &  & & \multicolumn{3}{c}{\textbf{Hitrate}} & \multicolumn{3}{c}{\textbf{nDCG}} &  & \multicolumn{3}{c}{\textbf{Hirate}} & \multicolumn{3}{c}{\textbf{nDCG}} \\ \cline{6-11} \cline{13-18} 
&  &  &  &  & \textbf{@10} & \textbf{@100} & \textbf{@300} & \textbf{@10} & \textbf{@100} & \textbf{@300} &  & \textbf{@10} & \textbf{@100} & \textbf{@300} & \textbf{@10} & \textbf{@100} & \textbf{@300} \\ \midrule
\multirow{7}{*}{$\mathbb{E}$} & DeepWalk & - & 4.19 & & 0.158 & 0.977 & 1.934 & 11.084 & 6.504 & 6.820 &  & 0.639 & 3.260 & 7.065 & 16.606 & 11.475 & 11.446 \\
 & LINE(1st) & - & 3.52 & & 0.115 & 0.875 & 1.597 & 10.372 & 6.182 & 6.049 &  & 0.478 & 2.675 & 5.703 & 13.717 & 10.567 & 10.070 \\
 & LINE(2nd) & - & 3.79 & & 0.163 & 0.938 & 2.028 & 11.381 & 6.454 & 6.836 &  & 0.660 & 3.864 & 7.545 & 16.753 & 11.608 & 11.695 \\
 & Node2Vec & - & 4.54 & & 0.145 & 0.966 & 1.917 & 11.195 & 6.529 & 6.842 &  & 0.631 & 3.227 & 6.892 & 15.966 & 11.168 & 11.438 \\
 & Metapath2Vec & - & 4.83 & & 0.129 & 0.981 & 1.967 & 11.301 & 7.307 & 6.536 &  & 0.682 & 3.939 & 7.564 & 16.499 & 12.275 & 11.927 \\
 & $\rm{\our{}^\mathbb{E}}$ & 92.664 & 4.43 & & 0.196 & 1.001 & 2.261 & 11.880 & 6.674 & 7.616 &  & 0.708 & 4.052 & 7.828 & 17.260 & 12.253 & 12.363 \\ \hline
\multirow{5}{*}{$\mathbb{C}$} & HyperML & 93.042 & 5.39 & & 0.225 & 1.323 & 2.509 & 12.488 & 7.609 & 8.142 &  & 0.972 & 4.231 & 7.595 & 20.419 & 14.619 & 14.060 \\
 & HGCN & 93.131 & 5.85 & & 0.245 & 1.474 & 2.623 & 12.598 & 7.927 & 8.417 &  & 0.999 & 4.363 & 7.689 & 20.631 & 14.847 & 14.418 \\
 & $\rm{\our{}^\mathbb{H}}$ & 93.158 & 5.57 & & 0.257 & 1.414 & 2.859 & 12.621 & 8.147 & 8.403 &  & 1.016 & 4.523 & 7.899 & 20.606 & 15.000 & 14.579 \\
 & $\rm{\our{}^\mathbb{S}}$ & 93.133 & 4.95 & & 0.239 & 1.488 & 2.590 & 12.534 & 8.003 & 8.410 &  & 0.954 & 4.166 & 7.370 & 20.08 & 14.451 & 14.092 \\
 & $\rm{\our{}^\mathbb{U}}$ & 93.244 & 5.32 & & 0.259 & 1.442 & 2.929 & 12.826 & 8.336 & 8.638 &  & 1.042 & 4.457 & 7.755 & 21.020 & 15.156 & 14.691 \\ \hline
\multirow{4}{*}{$\mathbb{M}$} & GIL & 93.412 & 6.77 & & 0.272 & 1.530 & 3.103 & 13.279 & 8.412 & 8.650 &  & 1.146 & 5.049 & 8.806 & 21.430 & 15.824 & 15.413 \\
 & $\rm{Product^{best}}$ & 93.526 & 5.64 & & 0.337 & 1.858 & 3.720 & 14.466 & 9.722 & 10.115 &  & 1.2221 & 5.175 & 8.895 & 22.158 & 16.283 & 15.844 \\
 & $\rm{M^2GNN}$ & 93.437 & 6.16 & & 0.295 & 1.676 & 3.405 & 13.768 & 8.985 & 9.214 &  & 1.187 & 5.112 & 8.742 & 21.658 & 15.938 & 15.565 \\
 & \our{} & \textbf{93.675} & 6.21 & & \textbf{0.342} & \textbf{1.877} & \textbf{3.736} & \textbf{14.659} & \textbf{9.79} & \textbf{10.141} &  & \textbf{1.252} & \textbf{5.261} & \textbf{9.023} & \textbf{22.205} & \textbf{16.415} & \textbf{15.959} \\ \cline{1-18}
 & Improv. & 1.09\% & 40.18\% & & 74.49\% & 87.51\% & 49.31\% & 23.39\% & 46.69\% & 33.15\% &  & 76.84\% & 29.84\% & 15.27\% & 28.65\% & 33.97\% & 29.09\% \\
\bottomrule
\end{tabular}
\end{table*}

The overall comparative results are shown in Table \ref{tab:overall_results}. The best results are in bold and \our{} always performs best. The last row shows the relative improvement compared to \our{} in Euclidean space. Without the changing of the model architecture, \our{} significantly outperform the vanilla Euclidean models, especially gains over 70 percent improvement in HR@10. It can be observed that:
\begin{itemize}
\item $\rm{\our{}^\mathbb{E}}$ exhibits superior performance compared with other Euclidean models. The reasons are mainly two-fold: i) The walk based methods such as DeepWalk and LINE are designed for isomorphic graphs and cannot distinguish different types of nodes and relationships; ii) $\rm{\our{}^\mathbb{E}}$ considers both hard and easy negative samples, which can achieve finer-grained discrimination than simple sampling strategy, especially on the large-scale graph.
\item From the perspective of subspace type, constant curvature methods $\mathbb{C}$ are always better than Euclidean methods $\mathbb{E}$. It can be seen that spherical space performs similarly to hyperbolic space, which indicates the hierarchical and cyclic structures both existing in our graph data. Notice that the unified space performs best, it aligns well with our intuition that we get better performance when we assign more flexibility in representation learning. Since different model layers are equipped with a different field view of data, the flexibility enables the curvature and spatial type to vary between layers, thus leading to a higher model capacity than other constant curvature spaces.
\item From the perspective of subspace number, mixed-curvature methods with multiple subspaces consistently outperform constant curvature methods. It's natural since the mixed-curvature manifold is able to represent multiple graph structures simultaneously, which preserves more topology information in our complex heterogeneous graph.
\item Throughout all methods, \our{} consistently outperforms all the competitive baselines over all the metrics. On the one hand,  \our{} adopts adaptive mixed-curvature space and jointly optimize subspaces, leading to higher capacity in characterizing complex graph structures. On the other hand, \our{} explicitly represents heterogeneous relations in different spaces, thus enhancing the utilization of type information.
\end{itemize}

Table \ref{tab:overall_results} also demonstrates the comparison of the training time for these models. Since the operations in curved spaces are more complex, the training time increases by 40 percent compared to Euclidean models in line with expectations. Due to the incremental learning, the time gap is relatively small and acceptable under the setting of daily system updates.

\subsection{RQ2: Ablation Study and Visualization}

\begin{table}[]
  \centering 
  \small
  \caption{Ablation analysis on our method. Each row indicates removing a module in the system model.}
\begin{tabular}{clccc}
\toprule
\multicolumn{2}{c}{\textbf{Model}} & \textbf{AUC} & \textbf{Hitrate@100} & \textbf{nDCG@100} \\ \hline
Full & AMCAD & \textbf{93.675} & \textbf{5.261} & \textbf{16.415} \\ \hline
\multirow{3}{*}{\begin{tabular}[c]{@{}c@{}}Node \\ 
Encoder\end{tabular}} & \textit{- mixed} & 93.244 & 4.457 & 15.156 \\
 & \textit{- curv} & 92.664 & 4.052 & 12.253 \\
 & \textit{- fusion} & 93.587 & 5.217 & 16.171 \\ \hline
\multirow{2}{*}{\begin{tabular}[c]{@{}c@{}}Edge \\ 
Scorer\end{tabular}} & \textit{- proj} & 93.205 & 4.413 & 14.995 \\
 & \textit{- comb} & 93.512 & 5.190 & 16.132 \\ 
\bottomrule
\end{tabular}
\label{tab:ablation_study}
\end{table}

\begin{figure}
  \centering
  \includegraphics[width=\linewidth]{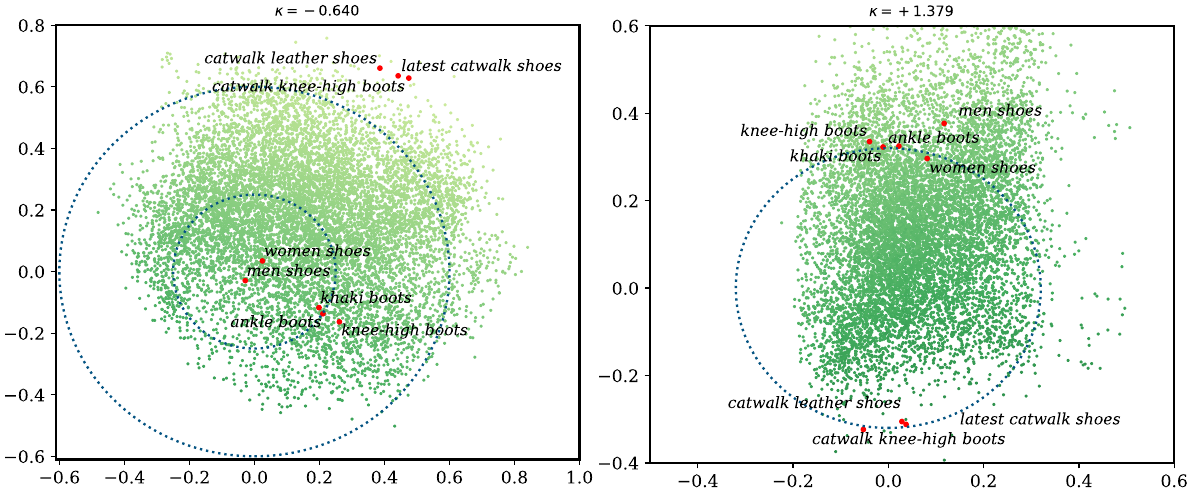}
  \caption{Illustration of the query embedding distribution in a mixed-curvature space composed of a hyperbolic space (left) and a spherical space (right).}
  \label{fig:vis_on_2dim}
\end{figure}

As discussed above, the performance gain of \our{}  is mainly due to the high capacity of model space and the utilization of heterogeneous information, corresponding to the \textit{node encoder} component and the \textit{edge scorer} component respectively. Ablation studies are conducted to quantify the gains of each component. Since we are more concerned with the model space than model architecture, ablation experiments on \textit{context encoding} are ignored.
\begin{itemize}
\item \textit{Inductive Learning.} Removing this module means that we discard the adaptive mixed-curvature space. We will discuss the importance of adaptivity in section \ref{sec:effects_of_parameters}. To verify the effectiveness of the adaptive mixed-curvature space, we build the model on a single unified space and Euclidean space, denoted by \textit{- mixed} and \textit{- curv} respectively.
\item \textit{Space Fusion.} Without this module, subspaces are optimized independently regardless of the common structure they represent. This branch is denoted by \textit{- fusion}.
\item \textit{Embedding Projection.} Since heterogeneous nodes are projected into different spaces according to their relations, we share the curvature of these edge spaces. In such a way, the graph is treated as a homogeneous graph. This change is denoted by \textit{- proj}.
\item \textit{Subspace-distance Combination.} Ignoring this module means that subspaces always have the same influence regardless of the local graph structure and type information. This experiment is represented as \textit{- comb}.
\end{itemize}

The ablation results are summarized in Table \ref{tab:ablation_study}. We observe that the curved space brings the greatest performance gain to the model. Mixed-curvature and heterogeneous projections are also important, which improve the space capacity for modeling complex heterogeneous graphs. The interactions and dynamic weights in subspaces slightly boost the results. Considering the enormous traffic in real industrial scenarios, even a tiny improvement brings substantial application benefits.

Besides these data analyses, we perform embedding visualization to give an intuitive illustration of the differences between our approach and the Euclidean approach. Toy representations are generated by setting both the number of subspaces and the dimension of the subspaces to be 2. The two learned subspaces of query nodes are hyperbolic and spherical shown in Fig.~\ref{fig:vis_on_2dim}. We randomly sample several queries under the same leaf category for illustration, which are marked as red points. As can be observed from the hyperbolic subspace in the left part of Fig.~\ref{fig:vis_on_2dim}, the hierarchical structure between these query nodes is well captured where the nodes with coarser semantic granularity are more likely to distribute near the space center. For instance, ``\textit{women shoes}" is closer to the origin than ``\textit{catwalk leather shoes}". Whereas, in the right part of Fig.~\ref{fig:tree_cyclic_structure_sample}, these query nodes exhibit cyclic distribution in the spherical subspace. The reason is that they belong to the same leaf category and are similar in semantics. Meanwhile, we found that the average weight of the hyperbolic subspace is greater than the spherical subspace, indicating that the hierarchical structure is more dominant in Q2Q relation.

\subsection{RQ3: Hyper-parameters Analysis} \label{sec:effects_of_parameters}

\begin{table}[]
\small
\centering
\setlength{\tabcolsep}{0.5em}
\caption{Evaluations on product method and our method. All models share the same settings except having different subspace types. Hyperbolic, Euclidean, spherical and unified subspaces are denoted by $\mathbb{H}$, $\mathbb{E}$, $\mathbb{S}$ and $\mathbb{U}$ respectively.}
\label{tab:effect_on_submanifold_type}
\begin{tabular}{ccccc}
\toprule
\textbf{Model} & \textbf{Subspace} & \textbf{Next AUC} & \textbf{Hitrate@100} & \textbf{nDCG@100} \\ \hline
\multirow{7}{*}{Product} 
 & $\mathbb{H} \times \mathbb{H}$ & 93.464 & 5.146 & 16.132 \\
 & $\mathbb{H} \times \mathbb{E}$ & 93.415 & 5.080 & 15.902 \\
 & $\mathbb{H} \times \mathbb{S}$ & 93.464 & 5.213 & 16.075 \\
 & $\mathbb{E} \times \mathbb{E}$ & 93.152 & 4.138 & 14.279 \\
 & $\mathbb{E} \times \mathbb{S}$ & 93.388 & 4.964 & 15.434 \\
 & $\mathbb{S} \times \mathbb{S}$ & 93.526 & 5.175 & 16.283 \\
 & $\mathbb{U} \times \mathbb{U}$ & 93.512 & 5.190 & 16.132 \\ \hline
 $\rm{\our{}}$ & $\mathbb{U} \times \mathbb{U}$ & \textbf{93.675} & \textbf{5.261} & \textbf{16.415} \\
\bottomrule
\end{tabular}
\end{table}

\begin{figure}
  \centering
  \includegraphics[width=0.8\linewidth]{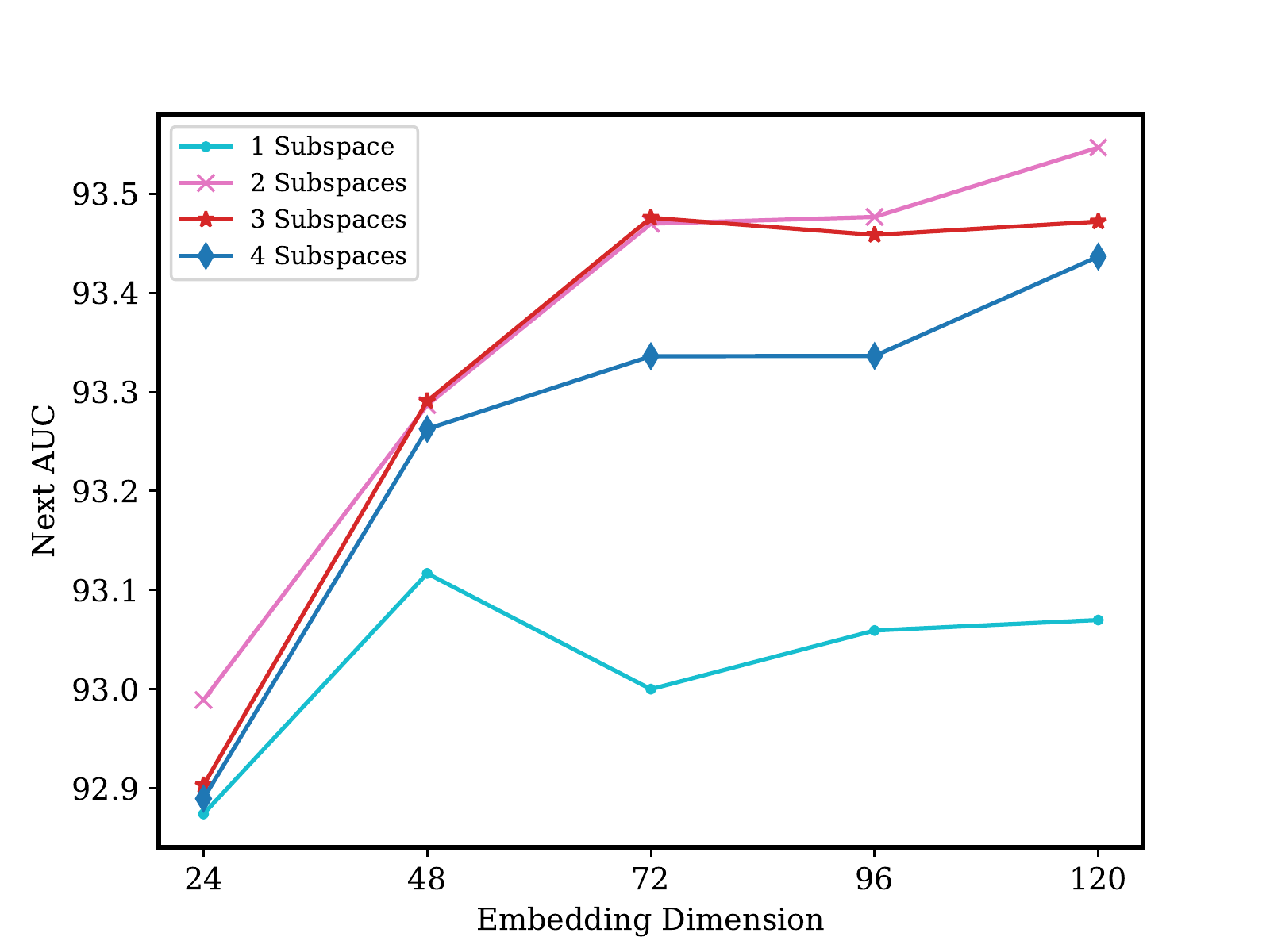}
  \caption{Next AUC score on our method with different subspace numbers and dimensions. For fair comparison, we compare models under the same total dimensions rather than the same subspace dimensions.}
  \label{fig:ablation_on_manifold_num}
\end{figure}

The mixed-curvature space contains three hyperparameters: the subspace type, curvature, and number. Due to the adaptivity of the unified manifold, only the number of subspaces need to be manually set. We first show the efficiency and accuracy of the adaptive property by comparing it with product space, then investigate the sensitivity of \our{} to the number of subspaces.

\subsubsection{Effects of adaptivity}

To evaluate the contribution of adaptive curvature, we compare \our{} with all possible combinations of product spaces. The results are shown in Table \ref{tab:effect_on_submanifold_type}. Our method performs best among these combinations due to the additional interactions between the subspaces. Besides that, the combination of pure unified subspaces is also better than most subspace combinations in product space. It proves that the unified manifold can always converge to the optimal space type and curvature.

\subsubsection{Effects of subspace numbers} \label{sec:submanifold_type}

Fig.~\ref{fig:ablation_on_manifold_num} shows the effect of subspace numbers under different embedding dimensions. Noticing that the embedding dimensions are the sum of all subspace dimensions for fair comparisons. The results show that a single subspace achieves its maximum performance ability at 48 dimensions and the space with two subspaces generally obtains the best performance. The reason for the worse performance on three or four subspaces is the low dimensionality of each subspace. From the trend of curve convergence, we believe more subspaces with sufficient dimensions lead to higher performance. 

\begin{table}[]
\small
  \centering
\caption{Training runtime comparisons of different graph size.}
\label{tab:training_runtime}
\begin{tabular}{ccccc}
\toprule
\textbf{Logs} & \textbf{\#Nodes} & \textbf{\#Edges} & \textbf{\#Iterations} & \textbf{Runtime (h)} \\ \midrule
1 hour & 2.7M & 0.18B & 0.19M & 0.5 \\
1 day & 106M & 5.3B & 5.75M & 6.2 \\
3 days & 190M & 16.1B & 17.48M & 17.3 \\
7 days & 300M & 30.8B & 30.07M & 35.0 \\ 
\bottomrule
\end{tabular}
\end{table}

\begin{figure}
  \centering
  \includegraphics[width=0.8\linewidth]{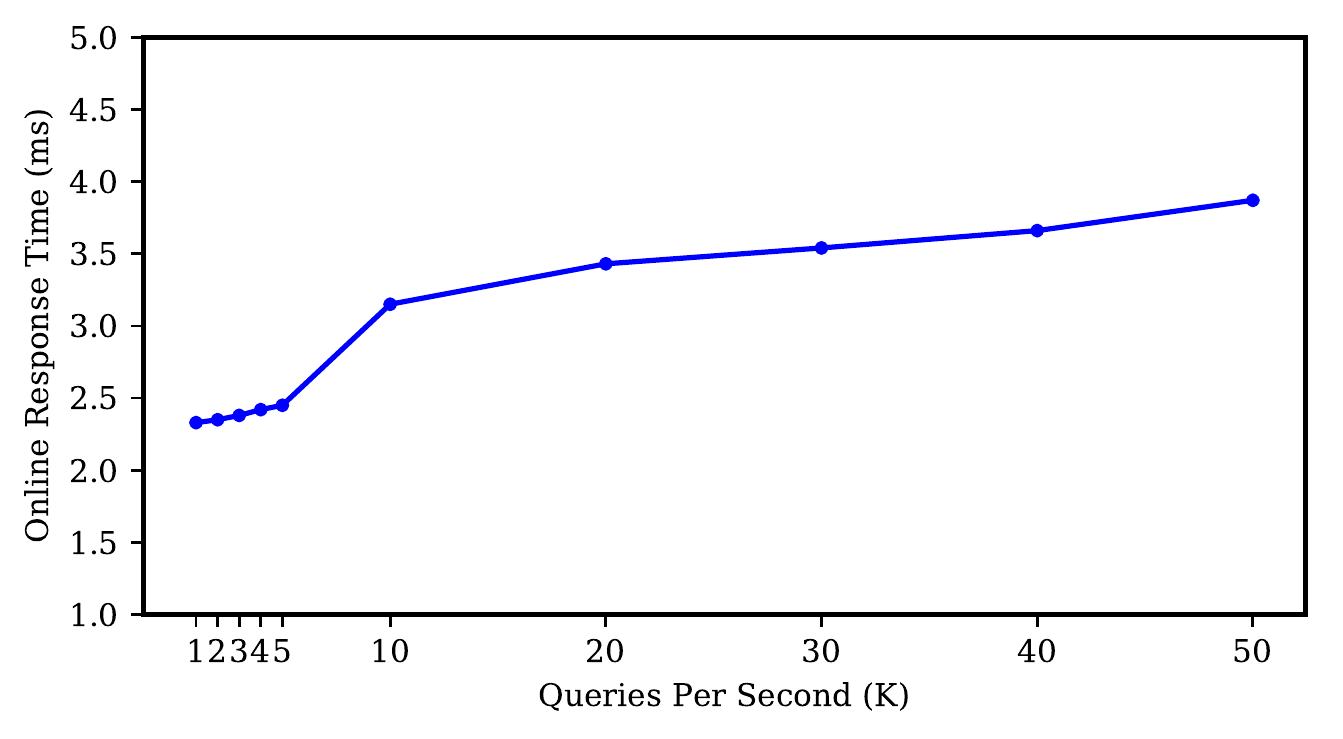}
  \caption{Statistics of online response time of ad retrieval with a different number of requests per second.}
  \label{fig:online_rt}
\end{figure}

\subsection{RQ4: System Efficiency}

Table \ref{tab:training_runtime} demonstrates the offline training runtime with the graph size varied. The graphs are constructed from user behaviors in different durations. We observe that the total training time raises near linearly with the number of graph edges, which indicates the high scalability of our system.

In the online serving stage, the two-layer inverted indexes are stored in the igraph engine. Fig.~\ref{fig:online_rt} illustrates the online response time of ad retrieval for different QPS (queries per second). The response time increases slowly and smoothly with QPS. With a tenfold increase in QPS, the response time has only doubled, indicating the efficiency of the serving architecture. The scalability of both offline training and online serving enables \our{} to process billions of streaming search traffic.

\subsection{RQ5: Online Results}

To evaluate the performance gain under real-time search traffic, we deploy \our{} in Taobao's sponsored search platform. We use click-through rate (CTR) and revenue per mille (RPM) to measure the performance: 
\begin{equation}
\begin{aligned}
    & \text{CTR}=\frac{\text{\#ad clicks}}{\text{\#impressions}} \\
    & \text{RPM}=\frac{\text{Ad revenue}}{\text{\#impressions}} \times 1000
\end{aligned}
\end{equation}
Note that there exist several ad retrieval channels in our sponsored search system, and one of them is our baseline $\rm{\our{}}^{\mathbb{E}}$. In our experiment, we only substitute this channel with our proposed method \our{} and keep other channels unchanged. By using a standard A/B testing configuration, we conduct the comparison on 4\% of the search request traffic of Taobao app.

Table \ref{tab:online_results_per_page} demonstrates the online results within 7 days, we can observe the overall CTR of \our{} method increases 0.5\%. This improvement verifies that our proposed method can touch more accurate ads matching the user requests. Moreover, the overall RPM increases 1.1\%, which indicates that \our{} method can also bring more revenue for our e-commerce platform. It is worth mentioning that the algorithms running on our platform have been highly optimized for many years, a lift rate larger than 1\% of RPM is viewed as a significant improvement. As for the single channel, the lift rates are more remarkable than the overall traffic, which are 1.5\% and 4.2\% for CTR and RPM respectively.

Table \ref{tab:online_results_per_page} also present the online results in different pages. It can be seen that our method achieves the largest improvement on the first page, and with the page number increasing, the improvement decreases. This phenomenon indicates that our method tends to retrieve more relevant and popular ads. Until now, \our{} is serving tens of billions of search requests per day on Taobao mobile app.

\begin{table}[]
\setlength{\tabcolsep}{0.5em}
\centering
\caption{Online result for A/B test within 7 days.} \small
\label{tab:online_results_per_page}
\begin{tabular}{ccccccc}
\toprule
\textbf{Metric} & \textbf{page 1} & \textbf{page 2} & \multicolumn{1}{l}{\textbf{page 3}} & \multicolumn{1}{l}{\textbf{page 4}} & \multicolumn{1}{l}{\textbf{page 5+}} & \multicolumn{1}{l}{\textbf{Overall}} \\ \midrule
CTR & +0.7\% & +0.4\% & +0.5\% & +0.3\% & +0.3\%  & +0.5\%\\ 
RPM & +1.5\% & +1.0\% & +0.7\% & +0.6\% & +0.7\%  & +1.1\% \\ 
\bottomrule
\end{tabular}
\end{table}

%% file: related.tex
\section{RELATED WORK}

A long-standing core problem in the ad retrieval system is how to measure the similarity between queries and ads. The most popular method is learning embeddings and matching functions to model real interaction data \cite{covington2016deep,ying2018graph,huang2020embedding}. Nevertheless, most works allocate embedding in Euclidean space, where large distortion exists when modeling complex industrial data \cite{sala2018representation,bachmann2020constant}.

Non-Euclidean space such as constant curvature space has gained attention in machine learning and network science due to its attractive properties for modeling data with certain types. In hyperbolic space with constant negative curvature, \cite{nickel2017poincare} proposed a generic symbol embedding approach and achieved significant performance for word embedding in the hierarchical WordNet dataset. Attractive performance also had been shown on a number of applications such as recommendation system \cite{vinh2020hyperml,chamberlain2019scalable}, question answering \cite{tay2018hyperbolic}, knowledge graph completion \cite{chami2020low}, graph classification \cite{liu2019hyperbolic} and similarity-based hierarchical clustering \cite{chami2020trees}. Besides, spherical space with positive curvature is proven to better capture cyclic structures. \cite{davidson2018hyperspherical,xu2018spherical} proposed the spherical variational auto-encoders and applied them in citation graphs and document modeling. \cite{meng2019spherical} developed a spherical generative model to jointly learn word and paragraph embeddings. Furthermore, the unified model of constant curvature space also had been applied to graph neural networks and outperformed the origin model \cite{bachmann2020constant}. Among these approaches, the sectional curvature is fixed, hence the space cannot capture multiple different local substructures simultaneously, which is ubiquitous in real-world data.

Embedding in mixed-curvature space is proposed to address the above problem. \cite{gu2018learning} proposed to learn the representation in the product of spaces, and formulated the mathematical operation like map functions and geodesic distance. \cite{skopek2019mixed} generalized the Euclidean variational auto-encoders to multiple curved latent spaces. \cite{zhu2020graph} combined Euclidean and hyperbolic geometries in a manner of dual feature interaction. \cite{wang2021mixed} applied mixed-manifold to model heterogeneous knowledge graphs. Though these methods are equipped with mixed-curvature, they either require enumerating subspaces types or treating subspaces equally and independently. Furthermore, these methods project nodes with different types into the same curved space, losing heterogeneity when representing industrial graph data.

%% file: conclusion.tex
\section{Conclusion}

In this paper, to model the complexity and heterogeneity of the query-item-ad interaction graph in our e-commerce sponsored search platform, we propose a fully adaptive mixed-curvature ad retrieval system named \our{}. The space curvatures and combinations are automatically learned in a data-driven way to model the complexity. Heterogeneous nodes and relations are mapped into different geometry spaces and evaluated attentively to model the heterogeneity. Extensive offline and online experiments have shown the efficiency and effectiveness of the system. We believe it will pave the new way to introduce non-Euclidean geometry to more industrial scenarios.